\documentclass[12pt]{iopart}

\usepackage{graphicx}
\usepackage{dcolumn}
\usepackage{bm}

\begin{document}

\title[Feasibility of QKD through DWDM network]{Feasibility of quantum key distribution through dense wavelength division multiplexing network}

\author{Bing Qi$^{1,2}$, Wen Zhu$^{2}$, Li Qian$^{1,2}$, and Hoi-Kwong Lo$^{1,2,3}$}

\address{$^1$Center for Quantum Information and Quantum Control, University
of Toronto, Toronto, Ontario, Canada\\
$^2$Dept. of Electrical and Computer Engineering, University of
Toronto, Toronto, Ontario, Canada\\
$^3$Dept. of Physics, University of Toronto, Toronto, Ontario,
Canada} \ead{bqi@physics.utoronto.ca}
\begin{abstract}
In this paper, we study the feasibility of conducting quantum key
distribution (QKD) together with classical communication through the
same optical fiber by employing
dense-wavelength-division-multiplexing (DWDM) technology at telecom
wavelength. The impact of the classical channels to the quantum
channel has been investigated for both QKD based on single photon
detection and QKD based on homodyne detection. Our studies show that
the latter can tolerate a much higher level of contamination from
the classical channels than the former. This is because the local
oscillator used in the homodyne detector acts as a ``mode selector''
which can suppress noise photons effectively. We have performed
simulations based on both the decoy BB84 QKD protocol and the
Gaussian modulated coherent state (GMCS) QKD protocol. While the
former cannot tolerate even one classical channel (with a power of
$0dBm$), the latter can be multiplexed with $38$ classical channels
($0dBm$ power per channel) and still has a secure distance around
$10km$. Preliminary experiment has been conducted based on a
$100MHz$ bandwidth homodyne detector.

\end{abstract}

\maketitle

\section{Introduction}

One important practical application of quantum information is
quantum key distribution (QKD), whose unconditional security is
based on the fundamental laws of quantum mechanics \cite{BB84,E91}.
Today most of the fiber-based QKD experiments are conducted through
dedicated ``dark'' fibers. As the applications of QKD have been
extended from point to point links to network configurations
\cite{Elliott02,Fernandez07,Peev09}, it would be much more appealing
if QKD can be conducted through the existing fiber network together
with classical communication signals. In this ``coexistence''
architecture, the quantum signal for QKD shares a common optical
fiber with unrelated classical signals \cite{Chapuran09}.

The coexistence architecture based on
wavelength-division-multiplexing (WDM) technology was proposed by
Townsend in the late nineties \cite{Townsend97} and has been studied
thereafter \cite{Chapuran09,Xia06,Xavier09,Peters09,Eraerds09}.
These studies show that the existence of strong classical traffic
could be detrimental to the QKD channel. This is because a classical
signal is typically many orders of magnitude stronger than the
quantum signal. Thus even a small crosstalk from a classical channel
could overwhelm normal QKD operation. One feasible way is to place
the quantum signal at the $1300nm$ ``original'' or O-band while
placing the classical signal at the $1550nm$ ``conventional'' or
C-band \cite{Chapuran09}. For such a large wavelength separation,
both the leakage of the classical signal and the anti-stokes Raman
scattering can be effectively suppressed to a tolerable level.

However, in many cases there are advantages to place both the
quantum and the classical signals in C-band. For example, the fiber
loss at C-band is significantly lower than that at O-band, so the
secure distance of QKD could be extended. Furthermore, this
architecture is more compatible with today's fiber network.

In \cite{Peters09}, a BB84 QKD system is multiplexed with classical
signals using C-band reconfigurable optical add drop multiplexer
(ROADM). In \cite{Eraerds09}, the authors multiplexed four classical
channels with one quantum channel using C-band $100GHz$
dense-wavelength-division-multiplexing (DWDM). This is a special
case where the four classical channels are used by the legitimate
users (Alice and Bob) for key distillation and encrypted
communication. The total power of the four classical channels is
only about $-22dBm$. By placing the quantum signal at a non-adjacent
channel of the classical signal, the maximum secure distance has
been shown to be around $40km$ \cite{Eraerds09}. These previous
studies suggest that with standard technology, it is still
infeasible to place both QKD signal and strong classical signal
($0dBm$) at C-band.

We remark that all the previous studies have been conducted in QKD
systems employing single photon detector (SPD), implementing, for
instance, the standard BB84 QKD protocol. Here, we will show that
QKD systems based on optical homodyne detection are inherently
robust against noise due to multiplexing. Intuitively, this is
because a strong local oscillator (LO) is used to beat with the weak
quantum signal during homodyne detection. Only photons in the same
spatiotemporal and polarization mode as the LO can be detected while
noise photons in different modes will be suppressed effectively.
Thus the LO acts as a ``mode selector'' \cite{Raymer95}. We remark
that in a recent free space QKD experiment, an optical homodyne
detector was employed to suppress ambient light \cite{Elser09}.

In this paper, we study quantitatively the amount of noise photon
added into the QKD channel in a typical DWDM setup. All the
theoretical simulations are carried out based on the performance of
commercial components. Two specific QKD protocols, namely, the decoy
state BB84 protocol \cite{decoy_theory,Ma05,Yi06} and the Gaussian
modulated coherent state (GMCS) QKD protocol \cite{GMCS_NATURE} have
been chosen for numerical simulations. The BB84 QKD protocol
\cite{BB84} is the first and the most well known QKD protocol. Its
unconditional security has been rigorously proved based on the laws
of quantum mechanics \cite{Security_BB84}, even when implemented on
practical setups with some imperfections, such as weak coherent
pulse, detector dark counts \cite{GLLP, Inamori07} and detector
efficiency mismatch \cite{Fung09}. The security of the GMCS QKD
protocol was first proven against individual attacks with direct
\cite{Grosshans02} or reverse \cite{GMCS_NATURE, Grosshans04}
reconciliation schemes. Security proofs were then given against
general collective attacks \cite{Navascues06, Garcia06, Lodewyck07}.
To date, three groups have independently claimed that they have
proved the unconditional security of GMCS QKD \cite{Security_GMCS}.

Our simulation results show that it is possible to multiplex the
GMCS QKD with a $0dBm$ classical channel using a C-band $100GHz$
DWDM without significantly reducing its performance. Even
multiplexed with $38$ \footnote{A typical 40-channel 100GHz DWDM
system with 1 channel for quantum communication and 38 nonadjacent
channels for classical communication.} classical channels ($0dBm$
power each channel), the GMCS QKD still has a secure distance around
$10km$. Preliminary experiment has been conducted based on a
$100MHz$ bandwidth homodyne detector.

We remark that although our studies are conducted in the GMCS QKD,
our conclusions should be applicable in other QKD protocols
employing homodyne detection such as the discrete modulated
continuous variable QKD protocol \cite{Hirano03}.

This paper is organized as follows: Section 2 contains theoretical
analysis of various noise sources in a generic quantum/classcial
signals coexistence scheme. In Section 3, We will quantify the
contribution from each noise source based on the performance of
commercial components. In Section 4, we will present simulation
results in both the decoy state BB84 protocol and the GMCS QKD
protocol. We will also present preliminary experimental results with
a 100MHz shot noise limited homodyne detector. Section 5 is a brief
conclusion.

\section{Theoretical analysis on noise contribution}

In a typical coexistence architecture based on DWDM, the noise
photons in quantum channel can be contributed by several sources
\cite{Chapuran09,Peters09,Eraerds09}, including the leakage photons
from the classical channels due to the finite isolation of the DWDM
components, the ``in-band'' noise photons generated in optical fiber
from nonlinear processes, such as four-wave mixing (FWM) and the
spontaneous Raman scattering, the in-band amplified spontaneous
emission (ASE) photons generated by optical amplifiers. Here,
in-band noise refers to noise photons within the spectral bandwidth
allocated to the quantum channel. In this section, we will quantify
the amount of noise photons contributed by each of the above sources
based on a typical DWDM configuration as shown in Fig.1. In Fig.1,
we assume that an erbium-doped fiber amplifier (EDFA) is employed to
boost the optical power of classical channels before multiplexing
with quantum channels. Furthermore, we assume that all the classical
channels are placed at wavelengths longer than that of the quantum
channels, since the spontaneous anti-Stokes Raman scattering (SASRS)
is typically weaker than the spontaneous Stokes Raman scattering.

In this paper, we assume that the eavesdropper (Eve) can control all
the classical channels and the EDFA (see Fig.1) but she cannot
access the multiplexer (MUX) and the demultiplexer (DEMUX) used for
multiplexing the quantum signals with classical signals. One special
example is that the classical signals are actually used by Alice and
Bob for authentication, error correction and privacy amplification
\cite{Eraerds09}. In the more general cases where the classical
channels are allocated to other users, Alice and Bob can place the
MUX and DEMUX in their local secure stations. Under the above
assumptions, the quantum signals sent by Alice are calibrated after
the MUX (point A in Fig.1). So the performance of the QKD system is
independent of the insertion loss of MUX. On Bob's side, the
insertion loss of the DEMUX can be treated as part of the loss in
Bob's detection system.

\begin{figure}[!t]\center
\resizebox{10cm}{!}{\includegraphics{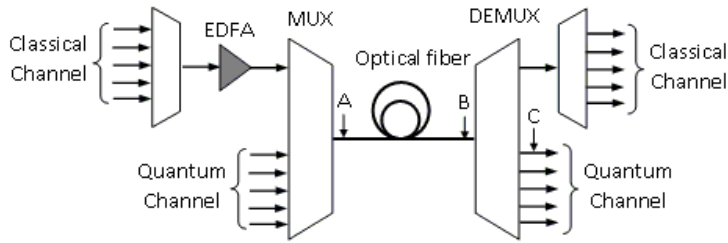}} \caption{A typical
scheme for multiplexing quantum channels with classical channels.
MUX-multiplexer; DEMUX-demultiplexer; EDFA-erbium-doped fiber
amplifier.}
\end{figure}

\subsection{Amplified spontaneous emissions (ASE) of EDFA}

It is well known that an ideal, noise-free amplifier cannot exist
\cite{Caves82}. In the case of an optical amplifier, the fundamental
noise originates from the spontaneous emission. The ASE from a
practical EDFA has a broad bandwidth on the order of tens of nm
which can be treated as a broadband noise source with a flat
spectral power density within the spectral bandwidth of the quantum
channel. We remark that a practical laser source also has a
broadband noise background, which can be modeled as ASE from a
virtual optical amplifier.

The average ASE photon number in one spatiotemporal mode is given by
\cite{Desurvire94}
\begin{equation}
\langle N_{ASE}\rangle=2n_{sp}(G-1)
\end{equation}
Here the factor $2$ accounts for the two orthogonal polarization
modes. $G$ is the gain of the EDFA, $n_{sp}\geq 1$ is the
spontaneous emission factor. If the spontaneous emission is the only
noise source (no excess noise), $n_{sp}=1$.

In practice, the excess noise of an EDFA is commonly quantified by
its noise figure ($NF$). In the unsaturated regime, $NF$ is related
to $n_{sp}$ by \cite{Desurvire94}
\begin{equation}
NF=\frac{1+2n_{sp}(G-1)}{G}
\end{equation}
In the high gain range ($G\gg1$), $NF\simeq 2n_{sp}$.

Typically, the ASE power is much lower than that of the classical
signal. However, its bandwidth is much broader and extends into the
quantum channel. Thus the ASE will contribute to in-band noise.
Fortunately, the MUX used at Alice's side functions as a bandpass
filter and can greatly suppress this in-band ASE noise. Given the
cross channel isolation of the MUX is $\xi_1$, the in-band ASE
photon number (per spatiotemporal mode) measured at the output of
the MUX (point A in Fig.1) is
\begin{equation}
\langle N_{ASE}^{(A)}\rangle=2\xi_{1}n_{sp}(G-1)
\end{equation}

Note, in this paper, we will not consider the ``out-of-band'' ASE
noise photons, since they are typically much weaker than the
classical signals themselves.

\subsection{Leakage from classical channel}

Although the classical signal has a different wavelength as the
quantum signal, a small fraction of the classical signal will leak
into the quantum channel due to the finite isolation of the DEMUX.
In a BB84 QKD system, this leakage will contribute to out-of-band
noise, which could be further reduced by using spectral filters at
the receiver's end. In a GMCS QKD system, this leakage contributes
noise photons in ``unmatched mode'' of the LO.

We define the power of the classical signal output from the
communication fiber as $P_{out}$ (measured at point B in Fig.1).
Given the isolation of the DEMUX $\xi_2$, the power of the leakage
signal received by Bob (measured at point C in Fig.1) is
$P_{leak}=\xi_{2}P_{out}$. The average leakage photon number per
second is
\begin{equation}
\langle N_{leak}^{(C)}\rangle=\frac{\xi_{2}P_{out}}{h\nu}
\end{equation}
Here $h$ is Planck's constant, and $\nu$ is the frequency of the
classical signal.

\subsection{Spontaneous anti-Stokes Raman scattering (SASRS)}

As the strong classical signals propagate along the optical fiber,
noise photons at different wavelength can be generated through
various nonlinear optical processes. If the wavelength of the noise
photons coincides with that of the quantum signal, they cannot be
filtered out at the receiver's end and will contribute to in-band
noise. It has been shown that SASRS is the dominant nonlinear
process when the quantum channel is placed at the shorter wavelength
of the classical channel \cite{Chapuran09,Eraerds09}.

The SASRS noise power within a bandwidth of $\Delta\lambda$
(measured at point B in Fig.1) is given by \cite{Chapuran09}
\begin{equation}
P_{SASRS}=P_{in}\beta{z}\eta_{ch}\Delta\lambda=P_{out}\beta{z}\Delta\lambda
\end{equation}
Here $\beta$ is the spontaneous Raman scattering coefficient,
$P_{in}$ (measured at point A in Fig.1) is the input power of the
classical signal, $z$ is the fiber length and $\eta_{ch}$ is the
transmittance of the optical fiber.

To estimate the noise photon number per spatiotemporal mode, we
first use the relation $\nu=c/\lambda$ to determine the total mode
number corresponding to a bandwidth of $\Delta\lambda$ and a time
window of $\Delta t=1s$ to be $N_{mode}=|\Delta\nu\Delta
t|=\frac{c}{\lambda^2}\Delta\lambda$. Here $c$ is the speed of light
in vacuum.

Given the insertion loss of the DEMUX is $\eta_{DMU}$, the in-band
SASRS photon number (per spatiotemporal mode) measured at the output
of the DEMUX (point C in Fig.1) can be calculated from (5):
\begin{equation}
\langle N_{SASRS}^{(C)}\rangle=\frac{P_{SASRS}}{h\nu
N_{mode}}\eta_{DMU}=\frac{\lambda^{3}}{hc^{2}}P_{out}\beta{z}\eta_{DMU}
\end{equation}
Again, in the derivation of (6), we have used the relation
$\nu=c/\lambda$.

\subsection{Four-wave mixing}

Four-wave mixing (FWM) is a third order nonlinear process generated
by the $\chi^{(3)}$ nonlinearity of the optical fiber when two or
more pumps exist. For FWM process to be efficient, phase-matching
condition is required. Although FWM could be the major noise source
at very short distance, it is much weaker than Raman Scattering for
a practical fiber length \cite{Peters09}. Furthermore, FWM can be
effectively suppressed by optimizing the channel configuration
\cite{Peters09,Eraerds09} or using polarization multiplexing
\cite{Peters09}. In this paper, we simply neglect FWM.

\section{Experimental characterization of various noise sources}

\subsection{The performance of commercial MUX/DEMUX}

Theoretical studies in Section 2 show that the noise level is
dependent on the performance of MUX/DEMUX: cross-channel isolation
and insertion loss.

We have tested two commercial C-band $100GHz$ MUX/DEMUX from JDSU
(model number: WD1508D1B). Each of them has 8 channels with a
channel separation of $0.8nm$ (or $100GHz$). The $3dB$ bandwidth of
each channel is around $0.6nm$ (or $75GHz$). Based on the
availability, a $1559.79nm$ fiber optic source module (ILX
Lightwave, 79800D) has been used as the laser source for the
classical channel. Thus, we allocate channel 2 of the MUX/DEMUX as
the classical channel and channel 8 as the quantum channel. The
isolation $\xi_1$ of MUX is determined by sending a calibrated laser
beam (with a wavelength of $\lambda_{8}$) into channel 2 and
measuring the optical power output from its common port. Similarly,
the isolation $\xi_2$ of the DEMUX is determined by sending a
calibrated laser beam (with a wavelength of $\lambda_{2}$) into its
common port and measuring the optical power output from its channel
8. In Table 1, we list the central wavelengths ($\lambda$),
insertion losses ($L$) and  the cross-channel isolations ($\xi$) of
two channels to be used in our experiments.

\begin{table}
\caption{JDSU C-band 100GHz MUX/DEMUX} \centering {
\begin{tabular}[t]{l  c  c  c  c  c}
\hline
Channel & $\lambda$(nm) & $L_{MUX}$(dB) & $L_{DEMUX}$(dB) & $\xi_1$(dB) & $\xi_2$(dB)\\
\hline
$2 (classical)$ & $1559.79$  & $2.61$ & $1.43$ & $-83$ & $-86$\\
$8 (quantum)$ & $1554.94$  & $0.97$ & $0.90$\\
\hline
\end{tabular}
}
\end{table}

\subsection{Noise photons contributed by amplified spontaneous
emissions of EDFA}

The noise contributed by an EDFA has been studied in Section 2.1.
However, the level of noise photon given in (3) is too low to be
measured with a conventional optical power meter directly. Instead,
we have performed an experiment based on a modified setup to test
the validity of (1). The modified experimental setup is shown in
Fig.2a. The EDFA is a commercial low noise fiber amplifier with a
$NF$ of 5.5dB (PriTel, LNHPFA-30-M). A tunable optical attenuator
(Att. in Fig.2) is used to simulate the transmission loss
experienced by the classical signal before it reaches the
multiplexer.

\begin{figure}[!t]\center
\resizebox{12cm}{!}{\includegraphics{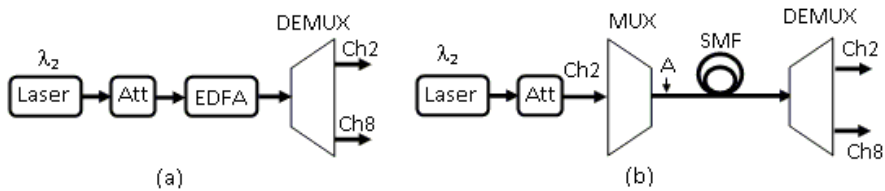}}
\caption{Experimental setups. Att-tunable optical attenuator;
SMF-single mode fiber.}
\end{figure}

Referring to Fig.2a, by setting the gain of EDFA to be $G=100$, we
have measured the noise power output from channel 8 to be
$P_{ASE}^{(Exp)}=-25dBm$. On the other hand, from (1) the ASE photon
is determined to be 351 per spatiotemporal mode. The energy of a
1550nm photon is $h\nu=1.28\times10^{-19}J$. So, the expected noise
power within $75GHz$ bandwidth (corresponding to 0.6nm) is
$351\times75\times10^9\times1.28\times10^{-19}J=-24.7dBm$.
Considering that the insertion loss of channel 8 is $0.9dB$, the
expected noise power $P_{ASE}^{(Theory)}=-25.6dBm$, which reasonably
matches with the experimental result $P_{ASE}^{(Exp)}$.

Next, we determine whether the background noise of the laser source
itself will make a significant contribution after being amplified by
the EDFA. We first determined the background noise of the laser
source. By using a tunable optical attenuator, the power of the
classical signal (input to the EDFA) was set to $-20dBm$. The laser
background noise at $1554.94nm$ ($\lambda_8$) within a $0.6nm$
wavelength range (corresponding to the channel bandwidth of
MUX/DEMUX) has been determined to be $-70dBm$. The number of in-band
laser noise photon in one spatiotemporal mode can be determined to
be around $0.01$. If the gain of EDFA is $G$, we expect that the
number of the amplified laser noise photon is around $0.01G$ per
spatiotemporal mode, which is significantly smaller than the number
of ASE noise photon (see (1)). In this paper, we will simply neglect
the contribution of the amplified laser noise.

\subsection{Spontaneous anti-Stokes Raman scattering generated in SMF28 fiber}

The SASRS noise photon number can be calculated from (6). The
spontaneous Raman scattering coefficient $\beta$ of standard SMF28
fiber has been determined using the experimental setup shown in
Fig.2b. The laser wavelength is $1559.79nm$ and the output power
after the MUX (point A in Fig.2b) was set to $4dBm$. We measured the
output power from channel 8 of the DEMUX for different fiber links:
a $20km$ SMF28 fiber spool and a $40km$ SMF28 fiber spool. Using
(5), the spontaneous Raman scattering coefficient has been
determined to be $2.85\times10^{-9}(km\cdot{nm})^{-1}$. Our result
matches with the results reported in \cite{Eraerds09}, which is
between $2\times10^{-9}\sim4\times10^{-9}(km\cdot{nm})^{-1}$
depending on the Raman shift.

In Section 4, we will calculate the secure key rates of both the
decoy state BB84 protocol and the GMCS QKD protocol in a typical
DWDM configuration as shown in Fig.1. The simulation parameters are
summarized in Table 2. To make our simulation results more
applicable, some parameters in Table 2, such as $NF$, $\eta_{MUX}$,
$\eta_{DMU}$, $\xi_1$ and $\xi_2$, have been chosen to be slightly
worse than the experimental values on purpose. Since the spontaneous
Raman scattering coefficient $\beta$ is wavelength dependent, we
have assumed the worst case of
$\beta=4\times10^{-9}(km\cdot{nm})^{-1}$ \cite{Eraerds09}. The
parameters of the GMCS QKD system are from \cite{Fossier09}.

\begin{table}
\caption{Simulation Parameters} \centering {
\begin{tabular}[t]{l l}
\hline
Parameter & Value  \\
\hline
NF (Noise figure of EDFA) & 4 (or 6dB)  \\
$\alpha$ (Fiber attenuation coefficient) & 0.21dB/km \\
$\beta$ (Spontaneous Raman scattering coefficient) & $4\times 10^{-9}/(km\cdot nm)$ \\
$\eta_{MUX}$ (Transmittance of MUX) & 0.71 (or 1.5dB loss) \\
$\eta_{DMU}$ (Transmittance of DEMUX) & 0.71 (or 1.5dB loss) \\
$\xi_1$ (Isolation of MUX) & $10^{-8}$ (or -80dB) \\
$\xi_2$ (Isolation of DEMUX) & $10^{-8}$ (or -80dB) \\
$\Delta{\nu}$ (3dB channel bandwidth) & 75GHz \\
$\Delta{t}$ (Gating window of SPD) & 1ns \\
$V_{A}$ (Modulation variance, GMCS) & 10 \\
$\eta_{Bob}$ (Transmittance of Bob's system, GMCS) & 0.6  \\
$\epsilon_{0}$ (GMCS parameter) & 0.01\\
$\upsilon_{el}$ (GMCS parameter) & 0.01 \\
$\gamma$ (Efficiency of reverse reconciliation algorithm) & 0.9 \\
\hline
\end{tabular}
}
\end{table}

\section{System comparison: single-photon detection scheme \emph{vs.} homodyne detection scheme}

\subsection{A single-photon detection based scheme: decoy state BB84 QKD with a weak coherent source}

Refer to Fig.1, in the BB84 QKD protocol, the quantum signal is at
the single photon level, so the crosstalk between the quantum
channels is negligible. We consider the simplest case where only one
classical channel (at a longer wavelength and non-adjacent channel)
is multiplexed with quantum channels.

In the BB84 QKD, single photon detectors are employed to detect
quantum signals. At telecom wavelength, InGaAs APDs working at the
gated Geiger mode are frequently used as SPDs. In this case, the
gating window of the SPD functions as a temporal filter, which can
reduce the effective noise photon number. For other non-gated
detection systems, this noise reduction can be achieved by
introducing an adjustable detection time window.

Given the channel bandwidth $\Delta{\nu}$ of MUX/DEMUX and the
gating window $\Delta{t}$ of the SPD, the total mode number of the
noise photons which can be detected by the SPD is
\begin{equation}
N_{Mod}=\Delta{\nu}\Delta{t}
\end{equation}

The number of noise photons arrived at Bob (point C in Fig.1) within
one gating window is
\begin{equation}
\langle N_{SPD}^{tot}\rangle=N_{Mod}\eta_{ch}\eta_{DMU}\langle
N_{ASE}^{(A)}\rangle+\langle
N_{leak}^{(C)}\rangle\Delta{t}+N_{Mod}\langle N_{SASRS}^{(C)}\rangle
\end{equation}
Here $\langle N_{ASE}^{(A)}\rangle$, $\langle N_{leak}^{(C)}\rangle$
and $\langle N_{SASRS}^{(C)}\rangle$ are determined from equations
(3), (4) and (6) respectively. At the RHS of (8), the superscripts
are used to refer to the location where noise photons are evaluated.

The gain of EDFA is adjusted with the channel transmittance
$\eta_{ch}$ to maintain a constant $P_{out}$ (point B in Fig.1) of
the classical channel. In this paper, we assume $G=100/\eta_{ch}$.

Using (3-8), we calculated the number of noise photons $\langle
N_{SPD}^{tot}\rangle$ as a function of the fiber length. We have
assumed that $P_{out}=0dBm$ and $G=100/\eta_{ch}$. Other simulation
parameters are summarized in Table 2. Fig.3 shows the simulation
results: at short distances, the main contribution of noise photons
is the leakage from the classical channel; while at long distances,
most of the noise photons are from SASRS. Noise contributed by the
EDFA is negligible.

\begin{figure}[!t]\center
\resizebox{10cm}{!}{\includegraphics{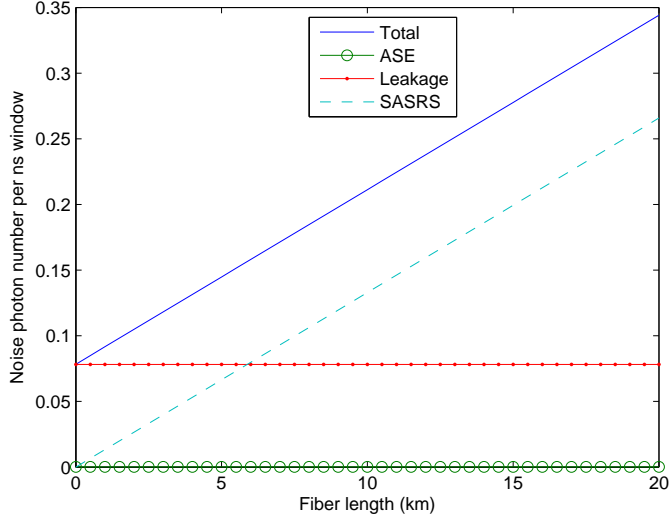}} \caption{The number
of noise photons arrived within 1ns gating window of SPD. The power
of the classical signal output from the communication fiber is
assumed to be $P_{out}=0dBm$ regardless of the fiber length. The
EDFA gain $G=100/\eta_{ch}$. Other simulation parameters are
summarized in Table 2.}
\end{figure}

In the BB84 QKD system, noise photons trigger random detection
events and contribute to quantum bit errors. This contribution can
be included in the total background count rate:
\begin{equation}
Y_{0}=Y_{0}^{0}+\eta_{Bob}\langle N_{SPD}^{tot}\rangle
\end{equation}
Here $Y_{0}^{0}$ is the background rate of the original QKD system.
The error rate of background counts is assumed to be $e_{0}=0.5$.

We assume that Alice and Bob perform perfect decoy state BB84
protocol with infinite decoy states. In the asymptotic limit of
infinitely many signals sent by Alice, the secure key rate (per
signal sent by Alice) is given by \cite{GLLP}
\begin{equation}
R=\frac{1}{2}[Q_{1}-f(E_{\mu})Q_{\mu}H_{2}(E_{\mu})-Q_{1}H_{2}(e_{1})].
\end{equation}
Here $\mu$ is the expected photon number of the signal state,
$Q_{\mu}$, $E_{\mu}$ are the gain and the overall quantum bit error
rate (QBER) of signal states, while $Q_{1}$, $e_{1}$ are the gain
and the QBER of single-photon components, $f\geq 1$ is the
inefficiency factor of the error correction algorithm. The estimated
values of the above parameters are given by \cite{Ma05}
\begin{equation}
Q_{\mu}=Y_{0}+1-e^{-\eta\mu}
\end{equation}
\begin{equation}
E_{\mu}=[e_{0}Y_{0}+e_{det}(1-e^{-\eta\mu})]/Q_{\mu}
\end{equation}
\begin{equation}
Q_{1}=(Y_{0}+\eta)\mu e^{-\mu}
\end{equation}
\begin{equation}
e_{1}=(e_{0}Y_{0}+e_{det}\eta)\mu e^{-\mu}/Q_{1}
\end{equation}
Here $e_{det}$ is the probability that a single photon hits the
wrong detector when Alice and Bob choose the same basis.
$\eta=\eta_{ch}\eta_{DMU}\eta_{Bob}$ is the overall efficiency,
where $\eta_{Bob}$ is the efficiency of Bob's system.

Given the noise photons shown in Fig.3, we calculated the secure key
rate of the decoy BB84 QKD system using (9-14). The simulation
parameters are chosen to be \cite{Eraerds09} $e_{det}=0.003$,
$Y_{0}^{0}=5\times10^{-6}$, $\eta_{Bob}=0.038$ and $f=1.22$. The
simulation results show that no secure key can be generated at any
distance. We remark that the simulation results are not sensitive to
the actual values of $e_{det}$ and $Y_{0}^{0}$ since the quantum bit
errors are mainly contributed by the noise photons due to
multiplexing.

\subsection{A homodyne detection based scheme: GMCS QKD}

The GMCS QKD has drawn a lot of attention for its potential high
secure key rate, especially at relatively short distances
\cite{GMCS_NATURE, Lodewyck07, Qi07, Fossier09}. In this protocol,
instead of performing single photon detection, Bob measures either
the phase quadrature or the amplitude quadrature of a weak coherent
state by using a homodyne detector. The strong LO used in homodyne
detection acts as a ``mode'' filter: only noise photons in the same
spatiotemporal and polarization mode as the LO can be detected,
while noise photons in unmatched modes will be effectively
suppressed. However, if the number of noise photons in unmatched
modes is comparable to the photon number of the LO, their
contributions cannot be neglected \cite{Raymer95}.

Refer to Fig.1, in the GMCS QKD protocol, the optical power of each
quantum channel is mainly determined by the operation rate and the
average photon number of LO. Given a 1MHz operation rate and a LO of
$10^8$ photons per pulse, the optical power of each quantum channel
is around $-19dBm$, which is significantly lower than the average
power of the classical signal ($0dBm$). Here, we simply neglect the
crosstalk between quantum channels. If a pair of QKD users can
access multiple quantum channels, the achievable secure key rate
will scale with the channel number. We further assume that all the
classical channels are at longer wavelengths of the QKD signals and
the total noise contributed by all the classical channels is simply
a summation of the noise contributed by each channel.

\subsubsection{Noise photons in matched mode}

Noise photons in the same spatiotemporal and polarization mode as
the LO are contributed by in-band ASE and SASRS. The number of noise
photons in matched mode arrived at Bob (point C in Fig.1) is
\begin{equation}
\langle N_{GMCS}^{in}\rangle=\frac{1}{2}m(\eta_{ch}\eta_{DMU}\langle
N_{ASE}^{(A)}\rangle+\langle N_{SASRS}^{(C)}\rangle)
\end{equation}
Here, the factor 1/2 is due to the polarization selection of the LO,
$m$ is the number of classical channel, $\langle
N_{ASE}^{(A)}\rangle$ and $\langle N_{SASRS}^{(C)}\rangle$ are
determined by (3) and (6).

Both the ASE and the SASRS can be modeled as output from a chaotic
source with Bose$\--$Einstein photon statistics \cite{Voss00,
Voss03}. For a chaotic source with an average photon number of
$\langle n\rangle$, the quadrature variances in shot noise unit are
given by \cite{Loudon00}:
\begin{equation}
(\Delta X)^2=(\Delta Y)^2=2\langle n\rangle+1
\end{equation}

Using (15) and (16), the ``excess noise'' contributed by noise
photons in matched mode is given by
\begin{equation}
\varepsilon_{in}=2\eta_{Bob}\langle N_{GMCS}^{in}\rangle
\end{equation}

\subsubsection{Noise photons in unmatched modes}

In the normal working condition, the pulse width of the LO is
significantly smaller than the integration time of the homodyne
detector. The latter is determined by the bandwith $\Delta f$ of the
homodyne detector and can be estimated by $\Delta
T=\frac{1}{2\pi\Delta{f}}$. Similar to (8), which gives the noise
photon number within one gating window $\Delta{t}$ of the SPD, the
number of noise photons in ``unmatched modes'' measured at Bob
(point C in Fig.1) within the time window of $\Delta{T}$ is given by
\begin{equation}
\langle N_{GMCS}^{out}\rangle=\frac{\Delta{T}}{\Delta{t}}\langle
N_{SPD}^{tot}\rangle
\end{equation}

Again, we model the noise photons in unmatched modes as the output
of a chaotic source. The photon number variance of a single mode
chaotic light is given by \cite{Loudon00}:
\begin{equation}
(\Delta{n})^2=\langle n\rangle^2+\langle n \rangle
\end{equation}
where $\langle n \rangle$ is the average photon number.

However, the integration time $\Delta T$ of the homodyne detector is
normally much larger than the coherence time $\tau_c$ of the noise
photon \cite{coherence-time}. Under this condition, the photon
number statistics follows Poisson distribution \cite{Milonni}. Thus
the ``excess noise'' contributed by noise photons in unmatched modes
is
\begin{equation}
\varepsilon_{out}=\eta_{Bob}\langle N_{GMCS}^{out}\rangle/\langle
n_{LO}\rangle
\end{equation}
Note in (20), we have added in a factor of $1/\langle n_{LO}\rangle$
to express $\varepsilon_{out}$ in shot noise unit, where $\langle
n_{LO}\rangle$ is the average photon number of the LO.

For example, if we assume that the bandwidth of the homodyne
detector is $\Delta{f}=1MHz$, then $\Delta{T}=0.16\mu s$. Since we
have assumed that the gating window of the SPD is $\Delta{t}=1ns$,
from (18), $\langle N_{GMCS}^{out}\rangle=160\times \langle
N_{SPD}^{tot}\rangle$. Based on the results in Fig.3, we expect that
the total number of noise photons in unmatched modes $\langle
N_{GMCS}^{out}\rangle$ is in the order of $10^2$. If we further
assume that $\langle n_{LO}\rangle=10^8$, then $\varepsilon_{out}$
is in the order of $10^{-6}$, which is negligible. In this paper, we
will neglect the contribution from photons in unmatched modes.

\subsubsection{Secure key rate of the GMCS QKD}

Under the ``realistic model'' \cite{GMCS_NATURE}, the secure key
rate (per signal sent by Alice) of the GMCS QKD with ``reverse
reconciliation'' protocol is given by \cite{Fossier09}
\begin{equation}
\Delta I=\gamma I_{AB}-\chi_{BE}
\end{equation}
where $\gamma$ is the efficiency of the reverse reconciliation
algorithm, and \footnote{To avoid confusion, some symbols are
different from the ones used in \cite{Fossier09}}
\begin{equation}
I_{AB} =\frac{1}{2}\log_2 [(V+\chi_{tot})/(1+\chi_{tot})]
\end{equation}
\begin{equation}
\chi_{BE}=\Theta(\frac{\sigma_1-1}{2})+\Theta(\frac{\sigma_2-1}{2})-\Theta(\frac{\sigma_3-1}{2})-\Theta(\frac{\sigma_4-1}{2})
\end{equation}
with $\Theta(x)=(x+1)log_2(x+1)-xlog_2x$;
$\chi_{tot}=\chi_{line}+\chi_{hom}/\eta_{ch}$;
$\chi_{line}=1/\eta_{ch}-1+\epsilon$;
$\chi_{hom}=(1+\upsilon_{el})/\eta'-1$;
$\sigma^{2}_{1,2}=\frac{1}{2}(A\pm\sqrt{A^2-4B)}$;
$\sigma^{2}_{3,4}=\frac{1}{2}(C\pm\sqrt{C^2-4D)}$;
$A=V^2(1-2\eta_{ch})+2\eta_{ch}+\eta^2_{ch}(V+\chi_{line})^2$;
$B=\eta_{ch}^2(V\chi_{line}+1)^2$;
$C=\frac{V\sqrt{B}+\eta_{ch}(V+\chi_{line})+A\chi_{hom}}{\eta_{ch}(V+\chi_{tot})}$;
$D=\sqrt{B}\frac{V+\sqrt{B}\chi_{hom}}{\eta_{ch}(V+\chi_{tot})}$.
Here, $V=V_{A}+1$ is the quadrature variance of the coherent state
prepared by Alice. $\eta'=\eta_{DMU}\eta_{Bob}$ is the equivalent
efficiency of Bob's system. $\epsilon$ denotes noise contribution
from outside of Bob's system, which can be further separated into
two terms:
\begin{equation}
\epsilon=\epsilon_{0}+\frac{\varepsilon_{in}}{\eta_{ch}\eta_{DMU}\eta_{Bob}}
\end{equation}
Here $\epsilon_{0}$ is contributed by the original GMCS QKD system.
$\varepsilon_{in}$ is the excess noise due to multiplexing with the
classical channels and can be determined from (17). Note that
$\varepsilon_{in}$ is determined at Bob's side, while $\epsilon$ and
$\epsilon_{0}$ are referred to input.

We have performed numerical simulations using parameters in Table 2
and the results are shown in Fig.4. In Fig.4, the secure key rates
have been calculated under 5 different conditions: (1) No classical
signal; (2) One non-adjacent classical channel ($0dBm$); (3) One
adjacent classical channel ($0dBm$); (4) 38 non-adjacent classical
channels ($0dBm$ per channel); (5) One non-adjacent classical
channel ($0dBm$) with our 100MHz homodyne detector (see details in
Section 4.2.4). In all the above cases, the optical power is defined
at the output of the communication channel (point B in Fig.1). Note
under condition (3), we have assumed that the isolation of MUX/DEMUX
between adjacent channels is $-40dB$, which is achievable with
commercial products \footnote{For the specific device we have
tested, the isolation between adjacent channels varies from channel
to channel in the range of $-30dB$ to $-50dB$}. From Fig.4, it is
possible to multiplex the GMCS QKD with a $0dBm$ classical channel
without significantly reducing its performance. Even multiplexed
with 38 classical channels ($0dBm$ power each channel), the GMCS QKD
still has a secure distance around $10km$. We remark that Eq. (23)
was derived from the realistic model \cite{GMCS_NATURE}, where Eve
cannot take advantages of noise contributed by Bob's system
\cite{Realistic_model}. In the more conservative ``general model''
\cite{GMCS_NATURE}, where Eve can control losses and noise in Bob's
system, the secure distance is much shorter \cite{Qi07}. In
practice, the realistic model has been commonly used
\cite{Lodewyck07, Qi07, Fossier09}.

\begin{figure}[!t]\center
\resizebox{10cm}{!}{\includegraphics{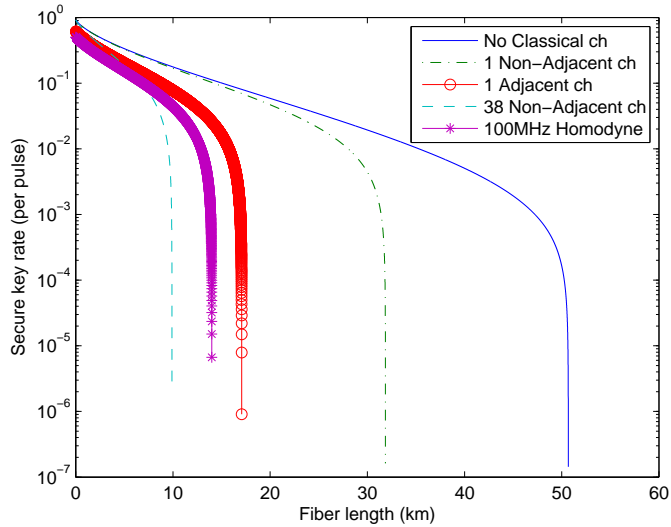}} \caption{The secure
key rate of the GMCS QKD under the ``realistic model''.  The secure
key rates have been calculated under 5 different conditions: (1) No
classical signal; (2) One non-adjacent classical channel ($0dBm$);
(3) One adjacent classical channel ($0dBm$); (4) 38 non-adjacent
classical channels ($0dBm$ per channel); (5) One non-adjacent
classical channel ($0dBm$) with our 100MHz homodyne detector (see
details in Section 4.2.4). Note under condition (3), we have assumed
that the isolation of MUX/DEMUX between adjacent channels is
$-40dB$. Other parameters are summarized in Table II.}
\end{figure}

\subsubsection{Preliminary experimental results with a 100MHz homodyne detector}

Recently, we have developed a $100MHz$ bandwidth shot-noise limited
optical homodyne detector \cite{Chi10}. With a LO photon number of
$4\times 10^8$, the electrical noise of the homodyne detector is
$10dB$ below the shot noise.

The noise photons output from the DWDM system (point C in Fig.1) has
been fed into the 100MHz homodyne detector. The fiber length used in
this experiment is $20km$. The variance of the output of the
homodyne detector has been measured as a function of the LO power
under three conditions: (1) No classical channel; (2) One classical
channel ($P_{out}=3.4dBm, G=100/\eta_{ch}$) is placed at channel 2,
channel 8 is used as the quantum channel; (3) One classical channel
($P_{out}=3.4dBm, G=100/\eta_{ch}$) is placed at channel 2, channel
1 is used as the quantum channel. Note under condition (3), the
quantum channel is placed at the adjacent channel of the classical
channel. Under both condition (2) and (3), the additional excess
noise due to multiplexing has been estimated from (17) to be less
than 0.01 (in shot noise unit). On the other hand, the measurement
uncertainty of our homodyne detection system on determining the
noise variance has been measured to be $\varsigma=0.024$ (in shot
noise unit) \footnote{The measurement uncertainty is defined as
three times of the standard deviation}. As shown in Fig.5, there is
no observable difference among the 3 measurement results. To
determine the magnitude of multiplexing noise more accurately, the
measurement accuracy of the homodyne detector has to be further
improved.

\begin{figure}[!t]\center
\resizebox{10cm}{!}{\includegraphics{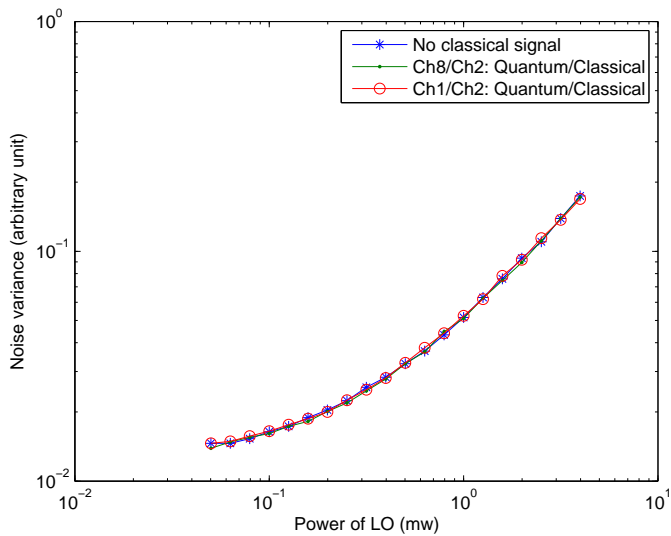}}
\caption{Experimental results with a 100MHz homodyne detector. The
total variance of the output of the homodyne detector has been
measured as a function of the LO power under three conditions: (1)
No classical channel; (2) One classical channel ($P_{out}=3.4dBm,
G=100/\eta_{ch}$) is placed at channel 2, channel 8 is used as the
quantum channel; (3) One classical channel ($P_{out}=3.4dBm,
G=100/\eta_{ch}$) is placed at channel 2, channel 1 is used as the
quantum channel. The fiber length is 20km. From the experimental
results, there is no observable difference among the 3 curves. }
\end{figure}

This raises up a question about how to apply the realistic model
using this homodyne detector, since Bob has to estimate the excess
noise from his system and the one from outside separately. To
determine the excess noise contribution from outside of Bob's system
($\epsilon$), one common way is to subtract the total observed noise
by the vacuum noise associated with the transmission efficiency and
the electrical noise ($\upsilon_{el}$) contributed by Bob's homodyne
detector. Obviously, the minimal $\epsilon$ resolvable by this
method is determined by the measurement uncertainty $\varsigma$ of
the homodyne detector \footnote{Assume that the transmission
efficiency (thus the vacuum noise) can be determined accurately}.

Note that the simulation results shown in Fig.4 (except the one
acquired under condition (5)) are based on parameters of the GMCS
QKD system from \cite{Fossier09}, where a low speed (1MHz bandwidth)
and low electronic noise (20dB below the shot noise) homodyne
detector was employed. So we have assumed that the measurement
uncertainty of the homodyne detector is small enough to resolve the
excess noise due to multiplexing. In our preliminary experiment,
however, the homodyne detector has a larger bandwidth (100MHz) but
also higher electronic noise (10dB below the shot noise). As a
conservative estimation of $\epsilon$, we could replace (24) by
\begin{equation}
\epsilon=\epsilon_{0}+\frac{\varepsilon_{in}}{\eta_{ch}\eta_{DMU}\eta_{Bob}}+\frac{\varsigma}{\eta_{ch}\eta_{DMU}\eta_{Bob}}
\end{equation}

Using (25), we have performed numerical simulations based on
parameters of the 100MHz homodyne detector: $\upsilon_{el}=0.1$ and
$\varsigma=0.024$. Other parameters are shown in Table 2.  We have
assumed that one non-adjacent classical channel ($0dBm$) is
multiplexed with the quantum channel. The simulation result is also
shown in Fig.4 (under condition (5)) where the secure distance is
about 14km. Note that the secure key rate under condition (5) is
significantly lower than that under condition (2). This is mainly
due to the larger measurement uncertainty of the 100MHz homodyne
detector. To apply the realistic model more efficiently, we may need
a more accurate way than the one given by (25) to estimate
$\epsilon$.

We would like to end this section with a few comments on the
realistic model adopted in the GMCS QKD. In all the security proofs
mentioned in Section 1, one underlying assumption is that both Alice
and Bob's QKD systems are fabricated by trusted vendors and these
devices are placed inside Alice and Bob's local secure stations
which cannot be accessed by Eve. So it might be reasonable to assume
that Eve cannot control the internal parameters of Bob's system (the
realistic model). However, to justify the above assumption in
practice, we may need to develop special techniques to estimate each
system parameter accurately without compromising the security of the
QKD system. As we have discussed in Section 1, to apply the security
proof of an idealized QKD protocol to a practical QKD system, all
the underlying assumptions and implementation details have to be
studied carefully.

\section{Conclusion}

In summary, we have studied the feasibility of conducting QKD
together with classical communication through the same fiber by
employing C-band DWDM technology. The impact of the classical
channel to the quantum channel has been investigated for both QKD
based on single photon detection and QKD based on homodyne
detection. Our studies show that the latter can tolerate a much
higher level of contamination from the classical channel than the
former. We have performed simulations based on both the decoy BB84
QKD protocol and the GMCS QKD protocol. With commercial DWDM
components, our simulation results show that it is possible to
multiplex the GMCS QKD with a $0dBm$ classical channel without
significantly reducing its performance. Even multiplexed with $38$
classical channels ($0dBm$ power each channel), the GMCS QKD still
has a secure distance around $10km$.

Although the LO is assumed to be a single mode coherent state in
this paper (which is not difficult to achieve in practice), it
doesn't have to be transform-limited.

The noise photons in the BB84 QKD system could be further reduced by
employing narrow spectral filter and temporal filter. For example,
in \cite{Eraerds09}, a $45pm$ spectral filter has been employed to
further cut off noise. In \cite{USpatent}, the author suggested to
use an additional optical gate to further suppress noise photons in
time domain. In principle, it is possible to selectively detect
photons in only one spatiotemporal mode by using an optimal
combination of spectral and temporal filters \cite{Qi07-2}. However,
in practice, both the spectral filter with extremely narrow
bandwidth and the temporal filter with extremely narrow time window
are difficult to fabricate and lossy as well as unstable (subject to
minute changes in temperature, pressure, etc.). Furthermore, Alice's
signal has to be transform-limited to pass through these filters
effectively. This also requires careful dispersion management in the
communication channel.

Financial support from CFI, CIPI, the CRC program, CIFAR, MITACS,
NSERC, OIT, and QuantumWorks is gratefully acknowledged.


\begin{thebibliography}{10}
\bibitem{coherence-time}
For example, if we assume that the bandwidth of the homodyne
detector is $\Delta{f}=1MHz$, then the integration time
$\Delta{T}=0.16\mu s$. The coherence time of the SASRS noise is
determied by the bandwidth $\Delta\nu$ of the DEMUX as
$\tau_c=\frac{1}{2\pi\Delta\nu}$. Using $\Delta\nu=75GHz$, we get
$\tau_c=2ps$. The coherence time of the classical signal depends on
both the laser source and the data rate. If we assume the data rate
of the classical channel is 10GHz, then the coherence time of the
classical signal is less than 100ps.
\bibitem{Realistic_model}
In the GMCS QKD, Alice and Bob estimate Eve's information from the
observed excess noise and other system parameters. A higher noise
indicates more information leakage and thus a lower secure key rate.
In practice, it may be reasonable to assume that Eve cannot control
devices inside Bob's system. Under this ``realistic model'', noise
inside and outside of Bob's system are treated differently: while
part of the excess noise (e.g., due to imperfections outside of
Bob's system) might originate from Eve's attack, the noise
contributed by Bob's devices is an intrinsic parameter of the QKD
system of which Eve has no control. Comparing with the more
conservative ``general model'' where Eve can control losses and
noise in Bob's system, the realistic model allows Alice and Bob to
acquire a tighter bound on Eve's information and thus yields a
higher secure key rate.
\end{thebibliography}
\end{document}